\documentstyle[twocolumn,epsf]{jpsj}
\def\NY{\nonumber \\}

\def\ct{\tilde{c}}
\def\at{\tilde{\alpha}}
\def\acs{\at_s / \ct_s}
\def\ach{\at_h / \ct_h}

\def\d{{\rm d}}
\def\i{{\rm i}}
\def\e{{\rm e}}

\def\r{\mbox{\boldmath $r$}}

\def\a{\mbox{\boldmath $a$}}

\def\runtitle{
Cooper Pair Formation in U(1) Gauge Theory 
of High Temperature Superconductivity
}
\def\runauthor{Atsuya {\sc Kumagai}, Masahiko {\sc Hayashi} 
and Hiromichi {\sc Ebisawa}}

\title
{
Cooper Pair Formation in U(1) Gauge Theory \\
of High Temperature Superconductivity
} 

\author
{ 
Atsuya {\sc Kumagai} \footnote{E-mail: kumagai@cmt.is.tohoku.ac.jp},
Masahiko {\sc Hayashi} 
and Hiromichi {\sc Ebisawa} 
}

\inst
{
Graduate School of Information Sciences, Tohoku University, Sendai 980-8579
}

\recdate
{
\today
}

\abst
{
We study the two-dimensional spin-charge separated Ginzburg-Landau theory 
containing U(1) gauge interactions as a semi-phenomenological model
describing fluctuating condensates in high temperature superconductivity.
Transforming the original GL action,
we abstract the effective action of Cooper pair.
Especially, we clarify how Cooper pair correlation evolves 
in the normal state from the point of view of spin-charge separation.
Furthermore, we point out how Cooper pair couples to gauge field
in a gauge-invariant way, stressing the insensitivity of Cooper pair
to infrared gauge field fluctuation.
}

\kword
{
high temperature superconductivity, RVB theory, gauge theory, 
Ginzburg-Landau theory, superconducting fluctuation
}

\begin{document}
\sloppy
\maketitle

The mechanism of high temperature superconductivity remains
still controversial after all the intensive efforts over a decade.
Among them, two-dimensional $t$-$J$ model has been studied as one of
the feasible microscopic models describing some anomalous properties in
the normal state. In the light of possible spin-charge separation
in the elementary excitations,
the phase diagram based on slave boson mean field approximation (MFA)
~\cite{SHF} shows fair correspondences with some experimental facts
and seems to indicate the origin of pseudogap.
On the other hand, it has also been pointed out~\cite{NL} that
only superconducting transition will remain if one takes the gauge field
coupling into account as the fluctuations around MFA, while that of spinon
and holon will vanish as artifacts accompanied with MFA and turn into
just crossovers. One of the functions of gauge field is 
to connect spin- and charge-subsystem
by eliminating the redundant degrees of freedom in the slave boson 
representation.
For instance, it appears in the electromagnetic responses of total system
as Ioffe-Larkin composition rule~\cite{IL}.
In addition, the gauge field is considered to cause
$T$-linear resistivity in the normal state by scattering 
charged particles(holons)~\cite{NL2}. Thus gauge field may play essential
roles in doped Mott insulators as a reflection of 
strong correlation among electrons in two dimensions.

When we consider the properties involving superconducting order
separately from the elementary excitations in the normal state,
what we need is a knowledge about the correlations of Cooper pair
as a composite of spin and charge, not that of each constituent.
In this Letter, we target how Cooper pair correlation evolves
in spin-charge separated description.
Starting from spin-charge separated Ginzburg-Landau (GL) theory 
containing U(1) gauge field,
we abstract the Cooper pair and its effective theory.
We shall also see how Cooper pair 'feels' gauge field and is affected
by its fluctuation, in comparison with the case of the separate constituents.
In the following, we set $\hbar = k_{\rm B} = c =1.$

We consider the circumstance where both pairing order parameters of spin 
and charge degrees of freedom are fluctuating with zero mean value,
which is a possible model describing the fluctuations of condensates
in the context of Resonating Valence Bond (RVB) theory.
For simplicity, we ignore the imaginary time dependence of the action.
Our starting point is two-dimensional, two-component Ginzburg-Landau action:
\begin{eqnarray}
& & S[\bar{s},s,\bar{h},h,\a] \NY
&=& \frac{1}{T} \int \d \r \left[ \right.
 \alpha_s |s(\r)|^2 + c_s|(-\i \nabla + \a(\r))s(\r)|^2 \NY
& & +\alpha_h |h(\r)|^2 + c_h|(-\i \nabla - \a(\r))h(\r)|^2 \NY
& &+\frac{u_s}{2} |s(\r)|^4 + \frac{u_h}{2} |h(\r)|^4 
+ u_c |s(\r)|^2|h(\r)|^2 \left. \right] , \label{GLsh}
\end{eqnarray}
thereby the partition function expressed as 
\begin{equation}
Z=\int D\bar{s} Ds D\bar{h} Dh Da \exp [-S[\bar{s},s,\bar{h},h,\a]].
\end{equation}
We take Coulomb gauge $\nabla \cdot \a(\r)=0$.
Similar model has been adopted in several papers
to investigate the properties concerning 
vortices in spin-charge separated systems~\cite{NL,FT}.
Complex scalar fields $s(\r)$ and $h(\r)$ represent the pairs of spinon and 
antiholon, respectively.
Although either of them should couple to the electromagnetic 
field, we set it aside for the present.
This action has a symmetry under the local U(1) gauge transformation
\begin{eqnarray}
& & s(\r) \rightarrow s(\r)\e^{\i \theta(\r)},
 \  h(\r) \rightarrow h(\r)\e^{-\i \theta(\r)}, \NY
& & \a(\r) \rightarrow \a(\r)-\nabla \theta(\r). \label{gt}
\end{eqnarray}
Each component couples to common U(1) gauge field $\a(\r)$.
Although $\a(\r)$ should have its own kinetic term stemming
from polarizations of the normal components~\cite{NL}, 
we omitted it to concentrate our attention upon the dominant couplings to
the fluctuations of the condensates.
Due to such implicit kinetic term, we assume that the gauge coupling is 
already in the weak-coupling region so that the perturbative treatment 
is justified.
The stability of the system requires the condition $u_s u_h - u_c^2 > 0$.
The last term in the action(\ref{GLsh}) is related to the correlation of
fluctuations between the components and therefore negative $u_c$ 
can be considered to accelerate the formation of Cooper pairs.
In the following, we limit our considerations to the cases $u_c < 0$.

Before analyzing the properties involving Cooper pair, let us give
a brief overview of some aspects of the model(\ref{GLsh}). 
For a while, we can consider the situation as if each component is
fluctuating independently, deferring the considerations of
the gauge coupling. 
At certain temperature around $\alpha_i=0 \ (i=s,h)$, 
each component is expected to go through a crossover from Gaussian 
to $XY$-like fluctuation, establishing a finite amplitude without
spontaneous symmetry breakings. This amplitude is considered to
cause a gap in such elementary excitation as spinon
in the context of slave boson mean field theory~\cite{SHF}.
We can also consider Kosterlitz-Thouless (KT) transition~\cite{KT}
in each subsystem.
Due to finite gauge coupling, however, the independent two-component 
phase fluctuation proves to be 
just a fictitious vision which will never occur in the real system.
Especially in the strong-coupling limit, it has been shown in several 
papers~\cite{FT,Rod} that after integrating out the gauge field first
there will finally remain three physical degrees of freedom: 
amplitudes $|s|, |h|$ and phase sum $\arg (sh)$.
Namely, it is (quasi-)long-range order of the phase of Cooper pair 
that survives even in the strong-coupling limit.
That is why we need to construct the effective theory of Cooper pair.

To see the kinetics of Cooper pair in the original action(\ref{GLsh}), 
we introduce auxiliary field $\Delta (\r) \sim u_c s(\r) h(\r)$:
\begin{eqnarray}
& & \exp\left[-\frac{1}{T} \int \d \r u_c |s(\r)|^2|h(\r)|^2 \right] \NY
&=& \int D\bar{\Delta}D\Delta \exp \biggl[ -\frac{1}{T} \int \d \r
   [ u_c^{-1}\bar{\Delta}(\r)\Delta(\r) \NY
& & -\bar{\Delta}(\r) s(\r)h(\r)-\Delta(\r) \bar{s}(\r)\bar{h}(\r) ] \biggr].
\end{eqnarray}
Then we have an action $S[\bar{s},s,\bar{h},h,\bar{\Delta},\Delta, \a]$.
The gauge symmetry of $\bar{\Delta}(\r) s(\r)h(\r)$ and its complex conjugate 
implies that
Cooper pair $\Delta (\r)$ is not transformed by (\ref{gt}); $\Delta (\r)$
is 'neutral' and not coupled to gauge field $\a(\r)$ through covariant
derivatives. Such coupling would be possible to the electromagnetic field.
We shall later see that Cooper pair couples to gauge field $\a(\r)$
in an alternative way.
If we integrate out $\bar{s}(\r),s(\r)$ and $\bar{h}(\r),h(\r)$, we obtain
the effective action with respect to Cooper pair and gauge field
as $S[\bar{\Delta},\Delta,\a]$:
\begin{eqnarray}
& & S[\bar{\Delta},\Delta,\a] \NY
&=& \frac{1}{T} \int \d \r \left[ \right.
 \alpha |\Delta(\r)|^2 + c |\nabla \Delta(\r)|^2 
  + \frac{u}{2} |\Delta(\r)|^4 + \cdots \NY
  & & + \chi (\nabla \times \a(\r))^2 + \cdots   \NY
  & & + v |\Delta(\r)|^2 (\nabla \times \a(\r))^2 + \cdots \left. \right] .
        \label{GLca}
\end{eqnarray}

\begin{figure}
\begin{center}
\hspace{0cm}
\epsfxsize=5.5cm
\epsffile{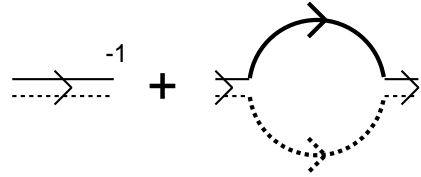}
\end{center}
\caption{
The formation of Cooper pair propagator. Double lines represent Cooper pair.
It acquires its own kinetics from spinon pair(solid lines) and 
antiholon pair(dashed lines). Thick lines mean the renormalized propagators
by each quartic term(see Fig.\ref{gl}).
}
\label{cooper}
\end{figure}
\begin{figure}
\begin{center}
\hspace{0cm}
\epsfxsize=7cm
\epsffile{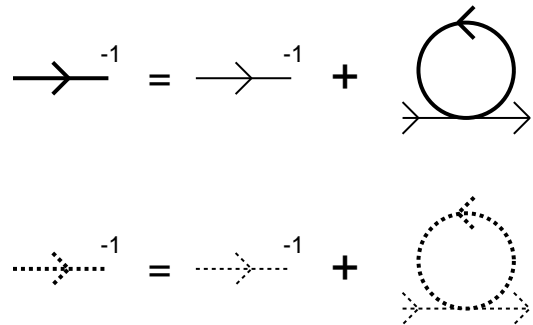}
\end{center}
\caption{
Renormalizations of propagators of spinon pair and antiholon pair
by each quartic term. 
}
\label{gl}
\end{figure}

\begin{figure}
\begin{center}
\hspace{0cm}
\epsfxsize=8cm
\epsffile{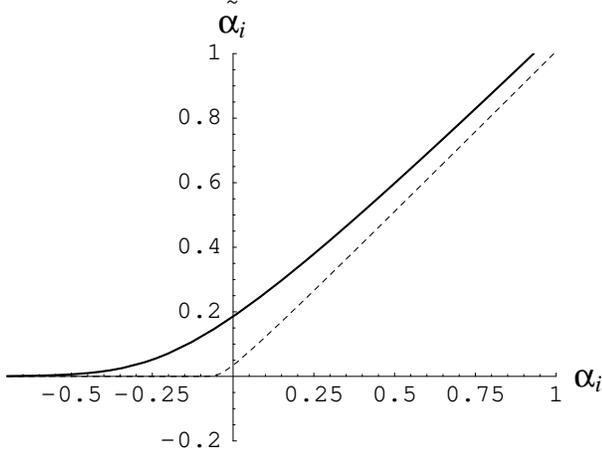}
\end{center}
\caption{ 
The relation between bare coefficient $\alpha_i$ and renormalized 
one $\at_i$. In this case holon pair(solid line) has much larger 
fluctuation than spinon pair(dashed line).
}
\label{aa}
\end{figure}

First we analyze the mechanism of the Cooper pair formation from
the viewpoint of its constituents.
We can derive the Cooper pair propagator from the processes
shown in Fig.\ref{cooper}. 
It is calculated by means of the propagators of spinon pair and antiholon pair
as if each component has independent kinetics described by GL theory,
deferring the integrating out gauge field; it is sufficient to renormalize
the propagator by each quartic term. 
Thus the coefficients $\alpha$ and $c$ are calculated as
\begin{equation}
 \alpha = - u_c^{-1} - \frac{T}{4 \pi} \frac{1}{\ct_s \ct_h} 
  \frac{\ln(\acs)-\ln(\ach)}{\acs - \ach},                   \label{mass}
\end{equation}
\begin{eqnarray}
 c = \frac{T}{2 \pi} \frac{1}{\ct_s \ct_h} & & \left[ 
  \frac{(\acs+\ach)(\ln (\acs) - \ln (\ach))}{2(\acs - \ach)^3}  \right. \NY
   & & \left. - \frac{1}{(\acs - \ach)^2} \right] ,
\end{eqnarray}
$\at_i$ and $\ct_i \  (i=s,h)$ meaning the renormalized coefficients 
by each quartic term.
Here we adopt the approximation as shown in Fig.\ref{gl}.
These diagrams mean that we take only
the renormalization of $\alpha_i$ by $u_i$ into account and neglect
any other renormalizations such as the renormalization of $u_i$ by $u_i$.
We shall later see that the renormalization of $\alpha_i$ is essential
for understanding the nature of Cooper pair fluctuation in the 'spin gap'
phase in the slave boson MFA~\cite{SHF}.
In this approximation we set $\ct_i = c_i$.
The renormalized coefficient $\at_i$ is determined by $\alpha_i$: 
\begin{equation}
  \at_i = \alpha_i + \frac{T u_i}{2 \pi c_s} 
    \ln \frac{c_i \Lambda^2 + \at_i}{\at_i} ,
\end{equation}
where ultraviolet cutoff $\Lambda$ is introduced in momentum space.
$\at_i$ has asymptotic forms
\begin{equation}
  \at_i \simeq \alpha_i \  (\alpha_i \rightarrow \infty),
\end{equation}
\begin{equation}
  \at_i \simeq \exp (2 \pi c_i \alpha_i / T u_i) 
          \  (\alpha_i \rightarrow -\infty).
\end{equation}
We give plots of the renormalized coefficients $\at_i$ as functions
of the bare coefficients $\alpha_i$ in Fig.\ref{aa}.
Here we set $\Lambda=1$, $c_s=c_h=1$, $u_s / 2 \pi=0.02$, $u_h / 2 \pi=0.2$,
which means the fluctuation of holon pair is much larger than that of 
spinon pair.
Around $\alpha_i=0$, the crossovers from Gaussian to $XY$-like fluctuations
are expected in the virtual subsystems described by GL theory.

\begin{figure}
\begin{center}
\hspace{0cm}
\epsfxsize=8cm
\epsffile{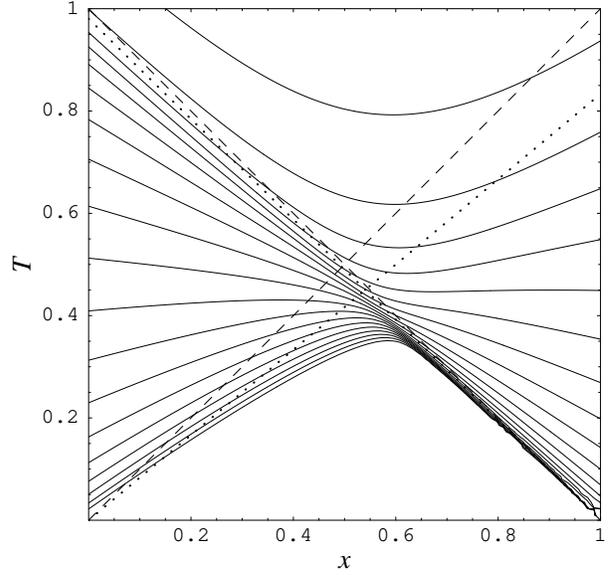}
\end{center}
\caption{A contour plot of GL coefficient $\alpha$ of Cooper pair as a
function of temperature and doping rate.
To see the wide-ranging variation, $\log (\alpha + u_c^{-1})$ is plotted
instead of $\alpha$ itself.
Bare transition temperatures $T_s^0 = 1-x, \ T_h^0 = x$ (dashed lines) and 
{\it fictitious} KT transition temperatures 
$T_i = T_i^0 ( 1 + u_i / \pi \alpha'_i c_i)^{-1} \ (i=s,h)$ (dotted lines)
are also plotted. 
}
\label{cc}
\end{figure}

Now we can see the evolution of the superconducting correlation
through GL coefficient $\alpha$ of Cooper pair.
We give a contour plot of $\alpha$ in Fig.\ref{cc}
based on eq.(\ref{mass}).
Here we set $\alpha_i = \alpha'_i (T-T_i^0) \  (i=s,h)$, 
$\alpha'_s = \alpha'_h = 1$, $T_s^0 = 1-x$, $T_h^0 = x$,
where $x$ can be regarded as a parameter corresponding to doping rate.
Bare transition temperatures $T_i^0 \ (i=s,h)$ and 
{\it fictitious} KT transition temperatures $T_i \ (i=s,h)$ are also plotted. 
$T_i$ are estimated as
\begin{equation}
  T_i = \frac{T_i^0}{1+u_i/\pi \alpha'_i c_i},
\end{equation}
following Halperin and Nelson~\cite{HN}.
With the temperature decreasing, $\alpha$ decreases monotonically.
We expect the superconducting (KT-)transition temperature $T_c$ is just below
such temperature as $\alpha=T-T_c^0=0$. 
Since $u_c$ accelerates Cooper pair formation,
the smaller is $|u_c|$, the lower are $T_c^0$ and $T_c$.
Combining Figs.\ref{aa} and \ref{cc},
one can see that the sharpness of the crossover is reflected in 
the sharpness of the drop in $\alpha$.
As a result, in the underdoped region, $\alpha$ has two sharp drops,
and in return, the superconducting fluctuation is larger than the overdoped 
region.
Even if $T_i$ and $T_i^0$ lose their meanings as characteristic temperatures
for the orderings of spinon pair or holon pair due to finite gauge 
couplings, they leave the traces on the evolution of Cooper pair correlation,
as explained in the following.
\begin{figure}
\begin{center}
\hspace{0cm}
\epsfxsize=7cm
\epsffile{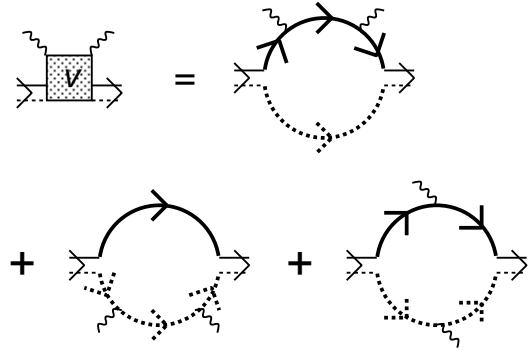}
\end{center}
\caption{
The lowest order coupling between Cooper pair and gauge field(wavy lines). 
}
\label{gauge}
\end{figure}

So far we have deferred the considerations of gauge field on the ground
that its effective coupling is comparatively weak.
In fact Cooper pair has a coupling to gauge field.
Such processes in the lowest order are depicted in Fig.\ref{gauge}.
One can confirm the cancellation of minimal couplings
between Cooper pair and gauge field
by letting momenta of gauge field zero.
Such cancellation in each order makes the destruction of ordering
in $s$ and $h$ by infrared gauge field fluctuation
invisible to Cooper pair.
Giving finite momentum to gauge field and expanding with respect to it, 
we evaluate the vertex $v |\Delta|^2 (\nabla \times \a(\r))^2$ as
\begin{eqnarray}
  v = \frac{T}{24 \pi} \frac{c_s / \at_s + c_h / \at_h}{\at_s \at_h}.
  \label{v}
\end{eqnarray}
This vertex should renormalize the Cooper pair propagator when gauge
field is integrated out. However,
because the gauge field never appears in the effective action 
$S[\bar{\Delta},\Delta,\a]$ without accompanying differential like 
$\nabla \times \a(\r)$,
Cooper pair is considered to be much less affected by infrared
gauge field fluctuations, in striking contrast to spinon pair or holon pair.

In concluding, we have abstracted the kinetics of Cooper pair
under the superconducting fluctuation
from the spin-charge separated Ginzburg-Landau theory with U(1) gauge field.
Our approach, which starts from the Gaussian fluctuations of two components
in high temperature and renormalizes it,
is complementary to that of Rodriguez~\cite{Rod},
where the amplitudes of the order parameters are fixed.

We first derived GL coefficients of Cooper pair
in general forms from the kinetics of its components.
Next we evaluated the renormalized coefficients $\at_i$, 
which proved to be essential for treating the region where
at least one of the components has $XY$-like fluctuation below $T_i^0$.
We assumed that the holon pair had much larger fluctuation than spinon pair;
it is the factor $u_i / \alpha'_i c_i$ that determines the magnitude
of the fluctuation of each order parameter.
In this case the sharp drop in $\alpha$ separates into two pieces
in underdoped region, located around $T_s$ and $T_h$,
while it concentrates around $T_s$ in overdoped region.
That leads to the enhancement of superconducting fluctuation
in underdoped region.

In terms of gauge field,
we implicitly assumed that its coupling is in the weak-coupling region 
on the ground that gauge field should already have its own kinetic term
due to the polarization of the normal component.
Because of the absence of the 'charge' of Cooper pair, it does not
couple to gauge field in a minimal way.
Instead we have pointed out the existence of the alternative couplings. 
We expect that the order of Cooper pair will not be destroyed critically
by gauge field fluctuation because gauge field always appears 
as its derivative,
unlike spinon pair or holon pair which directly connects to gauge field
through minimal coupling.

\section*{Acknowledgements}
We are grateful to Professor H. Fukuyama for encouraging discussions.
We are also indebted to Professor H. Kohno and Professor K. Kuboki
for fruitful discussions.

\end{document}